\documentstyle[mprocl]{article}
\begin{document}

\title{Is a `hadronic' shear current one of the sources in
  metric-affine gravity?} \author{Friedrich W. Hehl}
\address{Institute for Theoretical Physics, University of Cologne,
  D-50923 K\"oln, Germany} \author{Yuri N. Obukhov}
\address{Department of Theoretical Physics, Moscow State University,
  117234 Moscow, Russia}

\maketitle\abstracts{ The Minkowski space of special relativity can be
  understood as a flat 4-dimensional affine space enriched by a
  constant Minkowski metric.  If we gauge the general affine group and
  `superimpose' the metric, then we arrive at the metric-affine theory
  of gravity (MAG). The gravitational potentials are the spacetime
  coframe, the metric, and the linear connection. The material
  energy-momentum is coupled to the coframe (and the metric), a
  hypothetical hypermomentum current to the connection. The
  hypermomentum splits in a spin, a dilation, and a {\it shear} piece.
  We collect some evidence in favor of the existence of a material
  shear current in the context of Regge type trajectories of
  `hadronic' matter, thus supporting the link between particle physics
  and MAG.}

\vspace{-0.85cm}\section{Introduction}

``...the question whether this [spacetime] continuum is Euclidean or
structured according to the Riemannian scheme or still otherwise is a
genuine physical question which has to be answered by experience
rather than being a mere convention to be chosen on the basis of
expediency."\cite{ein}

Modern data convincingly demonstrate the validity of Riemannian
geometry on macroscopic scales. In the framework of Einstein's general
relativity theory (GR), mass (energy-momentum) of matter alone
determines the structure of spacetime at large distances.  However, at
high energies, the properties of matter are significantly different,
with additional spacetime related characteristics, such as spin and
scale charge coming into play. Correspondingly, one can expect that the 
geometric structure of spacetime on small distances may deviate from
Riemannian geometry.

\vspace{-0.25cm}\section{Formal structure of MAG} 

The structure of (flat) Minkowski space suggests the gauging of the
four-dimensional affine group
$T^4{\;{\rlap{$\supset$}\times}\;}GL(4,R)$, with the metric to be
superimposed. As gravitational potentials we find in this way the
coframe $\vartheta^\alpha$ (related to the subgroup of translations),
the linear connection $\Gamma_\alpha{}^\beta$ (linear subgroup), and
the metric $g_{\alpha\beta}$. The latter is not a gauge potential
proper, but rather takes values in the coset $GL(4,R)/SO(1,3)$ bundle
and is interpreted as a Higgs type field which emerges in the context
of symmetry breaking mechanism. ``If\dots the metric tensor is not a
fundamental physical entity but an order parameter of a quantum
condensate, then it is possible that at very high energies there is
some distorsion in nature''\cite{Finkelstein}, that is, nonmetricity
$Q_{\alpha\beta}:=-{\buildrel{\Gamma}\over{D}}g_{\alpha\beta}\neq 0$.
In GR, we have $Q_{\alpha\beta}=0$. The interplay between metric
$g_{\alpha\beta}$ (distance, angle) and connection
$\Gamma_\alpha{}^\beta$ (parallel displacement, inertial properties)
lays at the foundation of spacetime physics. The corresponding MAG
gauge field strengths are ($T^\alpha=$ torsion, $R_\alpha{}^\beta=$
curvature):
\begin{equation}
\bigl(Q_{\alpha\beta},\,T^\alpha,\,R_\alpha{}^\beta\bigr)
:=\bigl(-{\buildrel{\Gamma}\over{D}}g_{\alpha\beta},\,
{\buildrel{\Gamma}\over{D}}\vartheta^\alpha,\,
{\buildrel{\Gamma}\over{D}}\Gamma_\alpha {}^\beta\bigr)\,.
\end{equation}

The total action of the gravitational gauge fields and the minimally
coupled matter fields $\Psi$ reads\cite{PRs}
\begin{equation} 
W =\int [V(g_{\alpha\beta}, \vartheta^{\alpha}, Q_{\alpha\beta},
T^{\alpha}, R_{\alpha}{}^{\beta}) + L(g_{\alpha\beta},\vartheta^{\alpha},
\Psi , {\buildrel{\Gamma}\over{D}}\Psi)]\,.
\end{equation}
Matter currents arise from the matter Lagrangian $L$ as variational
derivatives:
\begin{equation}
\sigma^{\alpha\beta} := 2{{\delta L}\over{\delta g_{\alpha\beta}}},\quad 
\Sigma_{\alpha}:={{\delta L}\over{\delta\vartheta^{\alpha}}},\quad 
\Delta^{\alpha}{}_{\beta} := {{\delta L}\over{\delta\Gamma_{\alpha}
{}^{\beta}}}\,.
\end{equation}
Here $\sigma^{\alpha\beta}$ and $\Sigma_\alpha$ are metric and 
canonical energy-momentum current, respectively.

\vspace{-0.25cm}\section{Hypermomentum and shear current}

The hypermomentum current can be decomposed into 3 pieces:
\begin{eqnarray}
{\rm hypermomentum\> current\>\>}\Delta_{\alpha\beta}\>
&=&{\rm spin\>current\,\>}\Delta_{[\alpha\beta]}\\%
&\oplus&\>\,{\rm dilation\>}{\rm current\,\>}\Delta_{\gamma}{}^{\gamma}\>\\%
&\oplus&\>\,{\rm shear\>current\>\>}\Delta_{(\alpha\beta)}-
g_{\alpha\beta}\,\Delta_{\gamma}{}^{\gamma}/4\,.
\end{eqnarray}
Thus an independent (unconstrained) connection yields a new type of
{\it shear} current corresponding to the quotient $SL(4,R)/SO(1,3)$.
An $SL(3,R)$ current was originally proposed in the context of the
classification of sequences of hadrons\cite{Do4}. The Lie algebra of
the $SO(3)$ was extended by means of the five operators of the time
derivatives of the quadrupole moments of the `hadronic'
energy-momentum current. Later this procedure was generalized to the
$SL(4,R)$ and, eventually, to the general linear group $GL(4,R)$. For
the Dirac field, we can directly relate the time derivatives of the
quadrupole exitations to the (orbital) shear current\cite{Gr8},
\begin{equation}
{d\over dt}\int d^3x\,x^\alpha x^\beta\,\Sigma^{0\kappa}=
2\int d^3x\,x^{(\alpha}\Sigma^{\beta)\kappa}.
\end{equation}
(here $\Sigma^\alpha=\Sigma^{\kappa\alpha} \eta_\kappa$ and 
$\eta_\kappa:=\,^\star\vartheta_\kappa$). 

It is instructive to have a realistic model for matter with shear. For
continua with {\it microstructure}, such as in Mindlin's
model\cite{mindlin}, we have, besides the stress concept, that of an
hyperstress.  A 4-dimensional generalization of the Mindlin model
leads to the hyperfluid of Tresguerres and one of us\cite{Ob2}. The
basic variables are the {\it flow 3-form} $u$ of the hyperfluid, which
is related to its velocity vector field via
$u_{\alpha}:=e_{\alpha}\rfloor\,^\star u$, and three deformable {\it
  directors}. The hyperfluid is a continuous medium the elements of
which are characterized by the density of the classical `charges' of
the gauge group, that is, by the pair ($P_{\alpha}$, $J^{\alpha}_{\ 
  \beta}$) corresponding to the different generators of $GA(4,R)$.
Thus the material current 3-forms of the fluid are given by
$\Sigma_{\alpha}=uP_{\alpha}$ and $\Delta^{\alpha}_{\ \beta}=
uJ^{\alpha}_{\ \beta}\,.$ The variational principle\cite{Ob2} yields
explicitly,
\begin{eqnarray}
\Sigma_{\alpha}&=& \varepsilon uu_{\alpha} - p(\eta_{\alpha}-uu_{\alpha}) 
+ 2uu^{\beta}g_{\gamma [\alpha }\dot{J}^\gamma {}_{\beta ]}\,,\\%
\Delta^{\alpha}_{\ \beta} &=& u J^{\alpha}_{\ \beta},
\end{eqnarray}
where $\varepsilon=$ energy, $p=$ pressure, and 
$\dot{\Phi}:= -\,^\star {\buildrel{\Gamma}\over{D}}(u\,\Phi)$.

The dynamics of the hypermomentum is governed by the equations:
\begin{equation}
{\buildrel{\Gamma}\over{D}}\Delta^{\alpha}_{\ \beta} =
u^{\alpha}u_{\lambda}{\buildrel{\Gamma}\over{D}}
\Delta^{\lambda}_{\ \beta} + u_{\beta}u^{\lambda}{\buildrel{\Gamma}
\over{D}}\Delta^{\alpha}_{\ \lambda}.
\end{equation}
The hypermomentum density $J^{\alpha}_{\ \beta}$ satisfies a
generalized Frenkel type condition: $ J^{\alpha}_{\ 
  \beta}u^{\beta}=J^{\alpha}_{\ \beta}u_{\alpha}=0.$ This classical
description supposedly is a crude approximation to nature.

For the description of fundamental fermionic fields, we need to
generalize the special-relativistic Dirac spinor to MAG. ``The
dramatic way to general-relativize spinors is to add extra components
until we get a direct-sum representation that can be extended\dots But
in the case at hand, it takes an infinite-dimensional representation
of $GL_4$ to be double-valued. We would need infinitely many physical
partners for each spinor particle. A few brave people presently
explore this domain, especially Y.Ne'eman\dots''\cite{Finkelstein} In
this way one derives the so-called multi-spinors which carry infinite
many components. Technically they arise as unitary irreducible
representations of the covering group of $SL(4,R)$.  Typically they
are ordered in Regge type trajectories. Such matter should be used in
order to measure the nonmetricity of spacetime\cite{YF}. At low
energies, we expect a symmetry breakdown such that the affine group
reduces to the Poincar\'e group.

\vspace{-0.25cm}\section*{Acknowledgments}

This work has been supported by the Deutsche Forschungsgemeinschaft
(Bonn) project He-528/17-2.


\vspace{-0.25cm}\section*{References}
\begin{thebibliography}{99}

\bibitem{ein}
A. Einstein: {\it Geometrie und Erfahrung}. Sitzungsber. Preuss. Akad.
Wiss. (1921) 123-130; our translation.

\bibitem{Finkelstein} 
D.R. Finkelstein: {\it Quantum Relativity -- A Synthesis of the Ideas 
of Einstein and Heisenberg} (Springer, Berlin 1996) pp.\ 339 and 356. 

\bibitem{PRs} 
F.W. Hehl, J.D. McCrea, E.W.  Mielke, and Y. Ne'eman: {\it Metric-affine 
gauge theory of gravity: Field equations, Noether identities, world spinors, 
and breaking of dilation invariance}. Phys. Rep. {\bf 258} (1995) 1-171.

\bibitem{Do4}
Y. Dothan, M. Gell-Mann and Y. Ne'eman: {\it Series of hadron energy levels as 
representations of non-compact groups}. Phys. Lett. {\bf 17} (1965) 148-151.

\bibitem{Gr8} 
F. Gronwald and F.W. Hehl: {\it Stress and hyperstress as
fundamental concepts in continuum mechanics and in relativistic field
theory}, in: {\sl Advances in Modern Continuum Dynamics},
G. Ferrarese, ed. (Pitagora Editrice, Bologna 1993) pp. 1-32;
see also Los Alamos eprint archive gr-qc/9701054.
 
\bibitem{mindlin}
R.D. Mindlin: {\it Micro-structure in linear elasticity}. Arch. Rat.
Mech. Anal. {\bf 16} (1964) 51-78.

\bibitem{Ob2} Yu.N. Obukhov and R. Tresguerres: {\it Hyperfluid -- a
    model of classical matter with hypermomentum}. Phys. Lett. {\bf
    A184} (1993) 17-22.  

\bibitem{YF} Y. Ne'eman and F.W. Hehl: {\it Test matter in a
  spacetime with nonmetricity}.
  Class. Quant. Grav. {\bf 14} (1997) A251-A259.

\end{thebibliography}
\end{document}